\documentclass[iop,twocolappendix,appendixfloats]{emulateapj}
\usepackage{apjfonts,amsmath}
\usepackage{color}

\newcommand\beq{\begin{equation}}
\newcommand\eeq{\end{equation}}

\slugcomment{AJ, in press}  

\shorttitle{Triggering Nuclear Radio Activity}
\shortauthors{Lin, Huang, \& Chen}

\begin{document}


\title{An Analysis Framework for Understanding the Origin of Nuclear Activity in Low-Power Radio Galaxies}

\author{
Yen-Ting Lin\altaffilmark{1},
Hung-Jin Huang\altaffilmark{2},
and Yen-Chi Chen\altaffilmark{3}
}

\altaffiltext{1}{Institute of Astronomy and Astrophysics, Academia Sinica, Taipei 10617, Taiwan; ytl@asiaa.sinica.edu.tw}
\altaffiltext{2}{Department of Physics, Carnegie Mellon University, Pittsburgh, PA 15213, USA}
\altaffiltext{3}{Department of Statistics, University of Washington, Seattle, WA 98195, USA}

\begin{abstract}

Using large samples  containing nearly $2300$ active galaxies of low radio luminosity (1.4 GHz luminosity between $2\times 10^{23}$ and $3\times 10^{25}$\,W/Hz, essentially low-excitation radio galaxies) at $z\lesssim 0.3$,
we present a self-contained analysis of the dependence of the nuclear radio activity on both intrinsic and extrinsic properties of galaxies, with the goal of identifying the best predictors of the nuclear radio activity.  
While confirming the established result that stellar mass must play a key role in the triggering of radio activities, we point out that for central, most massive galaxies, 
the radio activity
also shows a strong dependence on halo mass, which is unlikely due to  enhanced interaction rates in denser regions in massive, cluster-scale halos.  
We thus further investigate the effects of various properties of the intracluster medium (ICM) in massive clusters on the radio activities,  employing two standard statistical tools, Principle Component Analysis and Logistic Regression.
It is found that  ICM entropy,  local cooling time, and pressure are the most effective in predicting the radio activity, pointing to the accretion of gas cooling out of a hot atmosphere to be the likely origin in triggering such activities in galaxies residing in massive dark matter halos. 
Our analysis framework enables us to logically discern the mechanisms responsible for the radio activity separately for central and satellite galaxies.

\end{abstract}


\section{Introduction} 
\label{sec:intro}

The cause of the nuclear radio activity in galaxies has long been a unsolved problem in astrophysics (e.g., \citealt{tadhunter16}).
With the advent of large scale surveys such as NRAO VLA Sky Survey (NVSS; \citealt{condon98}),  Faint Images of the Radio Sky at Twenty-Centimeters (FIRST; \citealt{becker95}), and Sloan Digital Sky Survey (SDSS; \citealt{york00}), it was clearly shown that the fraction of galaxies that are radio-loud (above certain luminosity threshold $P_{\rm th}$)\footnote{In the literature, there are different definitions for a galaxy to be radio-loud or radio-quiet/quiescent.  Throughout this work we simply use the radio luminosity to classify galaxies into these states.} is a strong function of stellar mass (e.g., \citealt{best05b,pasquali09}), and it is expected that the radio-loud phase is quite common in the course of the formation of massive galaxies (e.g., \citealt{heckman14}).  Indeed, in the current generation of galaxy formation models,  feedback from radio jets emanating from the central super massive black hole (SMBH) has been incorporated as an important mechanism for keeping massive galaxies ``red-and-dead'' 
(e.g., \citealt{croton06,bower06}; see also \citealt{mcnamara07}).
How  a very tight feedback loop can be maintained when the physical scales involved span 10 orders of magnitude remains a deep mystery \citep{fabian12}, however.

Occupying the most massive end of the galaxy population, the brightest cluster galaxies (BCGs) are found to exhibit the highest radio active fraction (RAF, defined to be the fraction of galaxies
selected with some specified stellar mass or optical luminosity range  with radio luminosity
$P\ge P_{\rm th}$, and, where necessary and possible, also certain specified halo mass range; RAF is about 30-40\% with $\log P_{\rm th}=23$ for BCGs in clusters at 1.4 GHz; e.g., \citealt{lin07,best07,vonderlinden07}; here $P$ is in unit of ${\rm W/Hz}$).  It is long observed that their nuclear radio activity cannot be solely attributed to the high stellar mass, however, as their proximity to the center of galaxy clusters  clearly plays important roles in triggering the radio active galactic nuclei (AGN).  For example, 
\citet{lin07} find that, compared to cluster galaxies of comparable stellar mass content (as traced by the near-IR luminosity), BCGs are more likely to be radio loud. 
It is also found that the spatial distribution of non-BCG member radio galaxies is highly concentrated towards the cluster center.  Both of these results indicate that 
galaxies in the central region of clusters have an enhancement of nuclear activity.
It is thus crucial to investigate both the effects of environments and internal properties of the galaxies on  triggering  SMBH activities.

It is suggested that the AGN population can be divided into two main categories, primarily based on the configuration of the central engine: a radiation dominated class (the so-called ``radiative mode''), and a mechanical power dominated class (the ``jet-mode'') \citep[e.g.,][]{ho08,heckman14}.  These roughly correspond to systems of high and low accretion rates onto the central SMBH, and therefore could be triggered by different physical mechanisms.
While radio-loud objects are found in both classes, the jet-launching mechanisms are likely different \citep[][and references therein]{heckman14,tadhunter16}.
To understand the phenomenology of radio-loud AGN, 
in addition to the central engine, one also needs to consider the properties and environments of the host galaxies  \citep[e.g.,][]{lin10b}.
In this paper, we shall focus on low-power (e.g., 1.4 GHz luminosity $\log P\le 25.5$) populations of radio AGN, which are overwhelmingly dominated by jet-mode/low-excitation systems \citep[e.g.,][]{best12,janssen12}, as they represent the majority of the radio AGN in the local Universe.

The goal of the present study is to seek  the key physical properties, both intrinsic and extrinsic to galaxies, that are most closely linked to the triggering mechanism of the radio activity in the nucleus. 
In addition to stellar mass and host halo mass, we shall also consider environmental factors such as local galaxy density and, for galaxies residing in clusters,  properties of the intracluster medium (ICM).
If such a link could be established with high statistical significance, we may be in a better position in identifying the most likely scenario among competing theoretical models (e.g., the precipitation model of \citealt{voit15}, the stimulated feedback model of \citealt{mcnamara16}) that aim to explain the radio AGN phenomenon.

There have been many studies attempting to discern the primary cause(s) of low-power nuclear radio activity.
For example, \citet{cavagnolo08} clearly demonstrate that radio emission is much more pronounced in BCGs when the  entropy of the ICM in the cluster center is lower than some threshold value.
Using a sample of 64 nearby clusters, \citet{mittal09} examine various correlations between the  radio luminosity of BCGs and cluster properties, finding a strong indication for the central cooling time of ICM to play an important role (see also \citealt{ineson15,mcnamara16}).
These results suggest the importance of gas cooling out of hot atmosphere/ICM surrounding the galaxies.
Extending the host dark matter halo mass range to include lower mass systems, \citet{pasquali09} examine the RAF as a function of both stellar and halo mass, finding a dominant stellar mass dependence over that on halo mass.
\citet{sabater13} find that, at fixed stellar mass, both dense environments and galaxy interactions enhance the likelihood of a galaxy being radio-loud.

For our investigation, we find it useful to separate galaxies into two classes: central and satellite.  Central galaxies are located close to the bottom of potential well of dark matter halos, and can grow in stellar mass via mergers with galaxies brought in by the dynamical friction, as well as by star formation due to (residual) cooling instability of the ICM.
Satellites, on the other hand, refer to all non-central galaxies in a galactic system; they are likely once central galaxies in their own dark matter halos before their halos get accreted/merged with the current halo.
In this picture, BCGs are central galaxies in massive clusters.
Given their different locations inside galactic systems,
it is plausible that the triggering mechanisms for central galaxies may be different from that for the satellites.

In this work we develop a framework of analysis that allows us to investigate the relative importance of various physical properties on the nuclear radio activity.  The analysis is carried out separately for central and satellite galaxies.
Using a large sample of radio galaxies associated with galactic systems that span a wide range in halo mass, we point out in Section~\ref{sec:halomass} that while for both central and satellite galaxies, stellar mass is a key predictor for radio activity, halo mass also plays an important role, particularly for the central galaxies.
Then in Section~\ref{sec:icm}, utilizing a cluster sample with detailed measurements of the ICM properties, we use robust statistical methods to point out that entropy, local cooling time, and pressure play the most significant role for the triggering of radio activities.
We conclude with a discussion on future prospects in Section~\ref{sec:disc}.

Throughout this paper we adopt a {\it WMAP5} \citep{komatsu09} $\Lambda$CDM cosmological model,
where $\Omega_m=0.26$, $\Omega_\Lambda=0.74$, $H_0=100h~{\rm km\,s^{-1}\,Mpc^{-1}}$ with $h=0.71$. 
All optical magnitudes are in the AB system.

\section{Importance of Halo Mass, Stellar Mass, and Local Galaxy Density} 
\label{sec:halomass}

In the first part of our analysis, we investigate the role of host galaxy stellar mass ($M_*$), dark matter halo mass ($M_h$), and local galaxy density ($\Sigma$) in triggering nuclear radio activity, using the radio galaxy (RG) population in groups and clusters found in SDSS.  After describing the construction of our RG sample,
we examine the dependence of RAF on these three physical properties, obtaining a qualitative picture from the global trends (Section~\ref{sec:sdssraf}); we then quantitatively compare the relative importance of these physical attributes via a {\it logistic regression} (LR) analysis (Section~\ref{sec:sdsslr}).

Our group and cluster sample is taken from the Data Release 7 (DR7) version of the group catalog of \citet{yang07}.  By assuming a one-to-one relationship between the halo mass and the total luminosity (or stellar mass) content of galaxy groups, \citet{yang07} are able to ``assign'' a halo mass to every galactic system they spectroscopically identify in SDSS (using essentially a matched-filter algorithm), down to single-galaxy systems.  For every galaxy group, they then designate the 
most massive 
galaxy closest to the geometric mean of member galaxy positions as the central galaxy; the rest are regarded as satellites.  
We adopt their central/satellite designation in this study.
We compute the stellar mass and absolute magnitudes of all member galaxies using the {\tt kcorrect} code \citep{blanton07}.

The RG sample used in this Section is based on two large RG catalogs: one is that of  \citet[][hereafter L10]{lin10b}, which covers the footprint of SDSS DR6 and is available in Table~\ref{tab:L10tab}, the other is taken from \citet{best12}, covering DR7.  Both studies cross match SDSS galaxy samples with 1.4 GHz radio source catalogs from NVSS and FIRST, largely following the methodology outlined in \citet{best05a}.  
To ensure the radio sources are powered (primarily) by an active nucleus, a combination of diagnostics is used to select RGs, including the BPT diagram \citep{baldwin81}, and the distributions of objects in the 4000\,\AA\ $vs.$ radio power, and  H$\alpha$ $vs.$ radio power planes \citep[see modifications discussed in][]{best12}.
We refer the reader to the original references for  detailed descriptions of the ways these catalogs are constructed; here we only point out two features that distinguish the L10 approach from the Best et al.~algorithm.  First, we start with a parent galaxy sample with $z\le 0.3$ and $M_r^{0.1}\le -21.27$ (i.e., more luminous than the characteristic magnitude in the galaxy luminosity function; \citealt{blanton03b}.  Here $M_r^{0.1}$ denotes the SDSS $r$-band shifted blueward by a factor of 1.1 in wavelength) from the New York University Value-Added Galaxy Catalog \citep{blanton05}.  Selecting RGs with a uniform absolute magnitude limit makes it straightforward to compute the RAF in volume-limited galaxy samples.
On the other hand, Best et al.~consider matches to a flux-limited radio source sample, and therefore some radio-loud galaxies may be missed.
Second, we have visually inspected all potential matches to 
improve the purity,
and to combine fluxes from distinct components for complex, extended sources.

\begin{figure}
	\includegraphics[width=0.8\columnwidth]{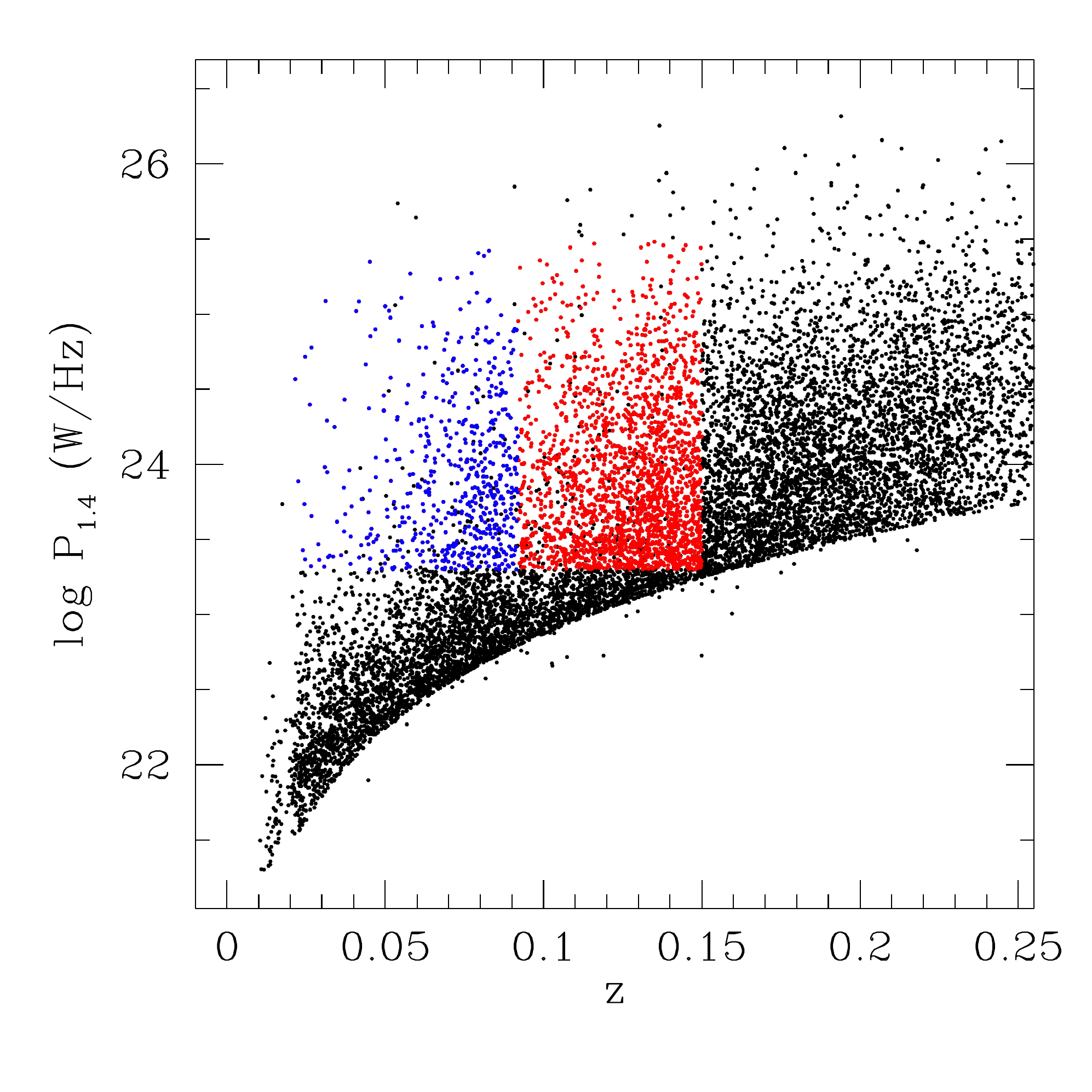}
    \caption{The distribution of our RG sample in the redshift--radio luminosity plane.  Our main galaxy sample, volume-limited to $z=0.15$, is represented by the union of red and blue-colored points, while the $z\le 0.092$ volume-limited sub-sample is shown as blue points (see Table~\ref{tab:sdssrg}).
}
\label{fig:lz}
\end{figure}

To construct the parent RG sample, we use all the objects from the L10 sample; in areas unique to DR7, we then use the \citet{best12} sample.
We further restrict ourselves to galaxies satisfying $z\le 0.15$, $M_r^{0.1}\le -21.57$, and
 $23.3 \le \log P_{1.4}\le 25.5$ ($P_{1.4}$ being the 1.4 GHz radio luminosity in W/Hz) 
to make the sample volume-limited in both optical and radio luminosities; these criteria also make our sample insensitive to the differences in the selection methods between L10 and \citet{best12}.  
Our selection essentially produces a low-excitation radio galaxy (LERG) sample, as only about 3\% of our RGs show strong enough [O\,{\sc iii}]  emission line to be classified as high-excitation radio galaxies (following the definition of \citealt{ching17}, that is, the line is detected at $\ge 3\sigma$, and has equivalent width $>5\,$\AA).  Our results do not change if we were to use the LERGs instead of a radio luminosity-selected  sample.
Our final sample consists of 2261 RGs 
(see Figure~\ref{fig:lz} for the distribution of our sample on the redshift--radio power plane).
Matching these to the 118638 members in the 97469 galactic systems from \citet{yang07}, we find that there are 1861  and 360 radio-loud central and satellite galaxies, respectively 
(see Table~\ref{tab:sdssrg}).
For halo mass we adopt the stellar mass ranking-based one from Yang's catalog (specifically, we use $M_h=M_{200b}$,  the mass enclosed in $r_{200b}$, within which the mean overdensity is 200 times the mean density of the Universe).

\begin{table}
	\centering
	\caption{RG samples in Yang et al.~groups}
	\label{tab:sdssrg}
	\begin{tabular}{lrr} 
		\hline
		main ($z\le 0.15$) sample\\
		\hline
		& central & satellite\\
		parent & 97469 & 21169 \\
		RG & 1861 & 360  \\
		\hline
		$z\le 0.092$ sub-sample\\
		\hline
		& central & satellite\\
		parent & 20529 & 4972 \\
		RG & 364 & 96 \\
		\hline
	\end{tabular}
\end{table}

\subsection{Global Trends of Radio Active Fraction}
\label{sec:sdssraf}

\begin{figure}
	\includegraphics[width=0.8\columnwidth]{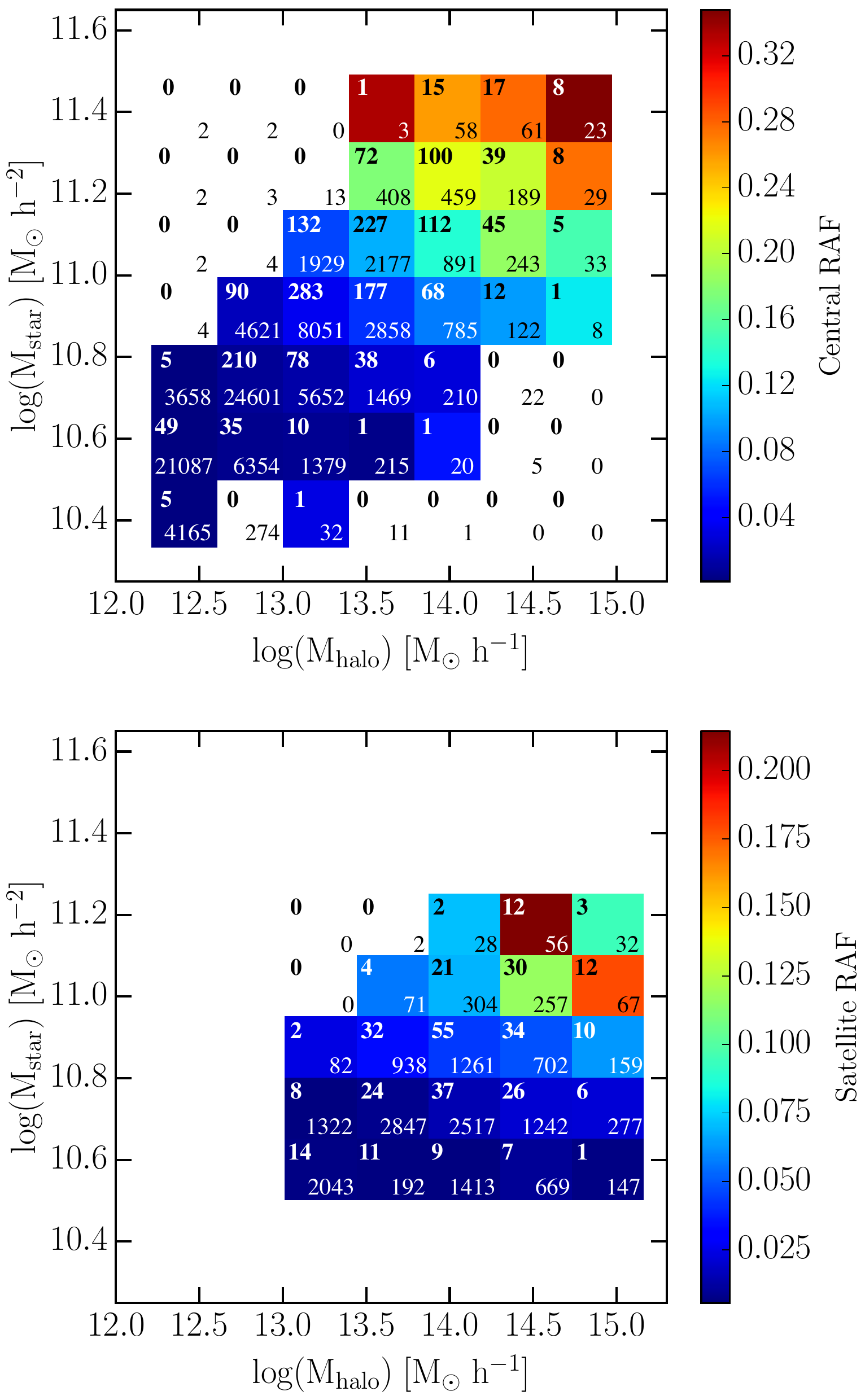}
    \caption{The dependence of RAF, defined as the ratio of number of RGs to that of all galaxies in a given $(M_*,M_h)$ bin, 
     on dark matter halo mass and stellar mass for central (top panel) and satellite (bottom panel) galaxies, based on our main sample.  It is apparent that the RAF depends on both halo mass and stellar mass for central galaxies; for satellites, the RAF is primarily dependent on stellar mass.
     In each $(M_h,M_*)$ bin, the number in the lower right  corner denotes the number of all galaxies, while the one in the upper left  corner is the number of RGs.
}
\label{fig:mhms}
\end{figure}

We show in Fig.~\ref{fig:mhms} the dependence of RAF $f(M_*,M_h)$ in the host galaxy stellar mass $vs.$ host group/cluster mass plane, for central (top) and satellite (bottom) galaxies.
Here the RAF is the ratio of number of RGs to that of all galaxies in a given $(M_*,M_h)$ bin. 
The widths of the two dimensional bins are chosen such that most of them contain at least 5 RGs, while small enough to reveal global trends with $M_*$ or $M_h$. 
Our results do not sensitively depend on the choice of bin widths.
It can be seen that the RAF dependence is different for centrals and satellites, in the sense that while satellite RAF is primarily a function of stellar mass (with only weak dependence on halo mass), both stellar mass and halo mass matter for centrals.
Our result is consistent with the findings of \citet{pasquali09}, who also study the stellar mass and halo mass dependence of RAF in groups and clusters identified by Yang et al. (although they have combined centrals and satellites in their analysis on RGs).
Both the group/cluster  and the RG samples we use are of much larger sizes compared to those used by \citet{pasquali09}. Our RG sample selection is also better defined
(in terms of radio flux limit, optical luminosity threshold, and visual inspection for completeness and purity).
Both of these factors make our results more statistically significant.

The trend we see in Fig.~\ref{fig:mhms} may be partially driven by the intrinsic correlation between stellar mass and halo mass for central galaxies (e.g., \citealt{yang07,mandelbaum16}).  To examine how much radio activity is caused by halo mass-dependent physical processes additional to the stellar mass--halo mass relation, at fixed stellar mass, 
we sum the RAF values over all halo mass bins, then normalize the RAF by that sum; in short, the normalized RAF is $\tilde{f}(M_h|M_*)=f(M_h,M_*)/\int f(M_h,M_*) dM_h$.

\begin{figure}
	\includegraphics[width=0.8\columnwidth]{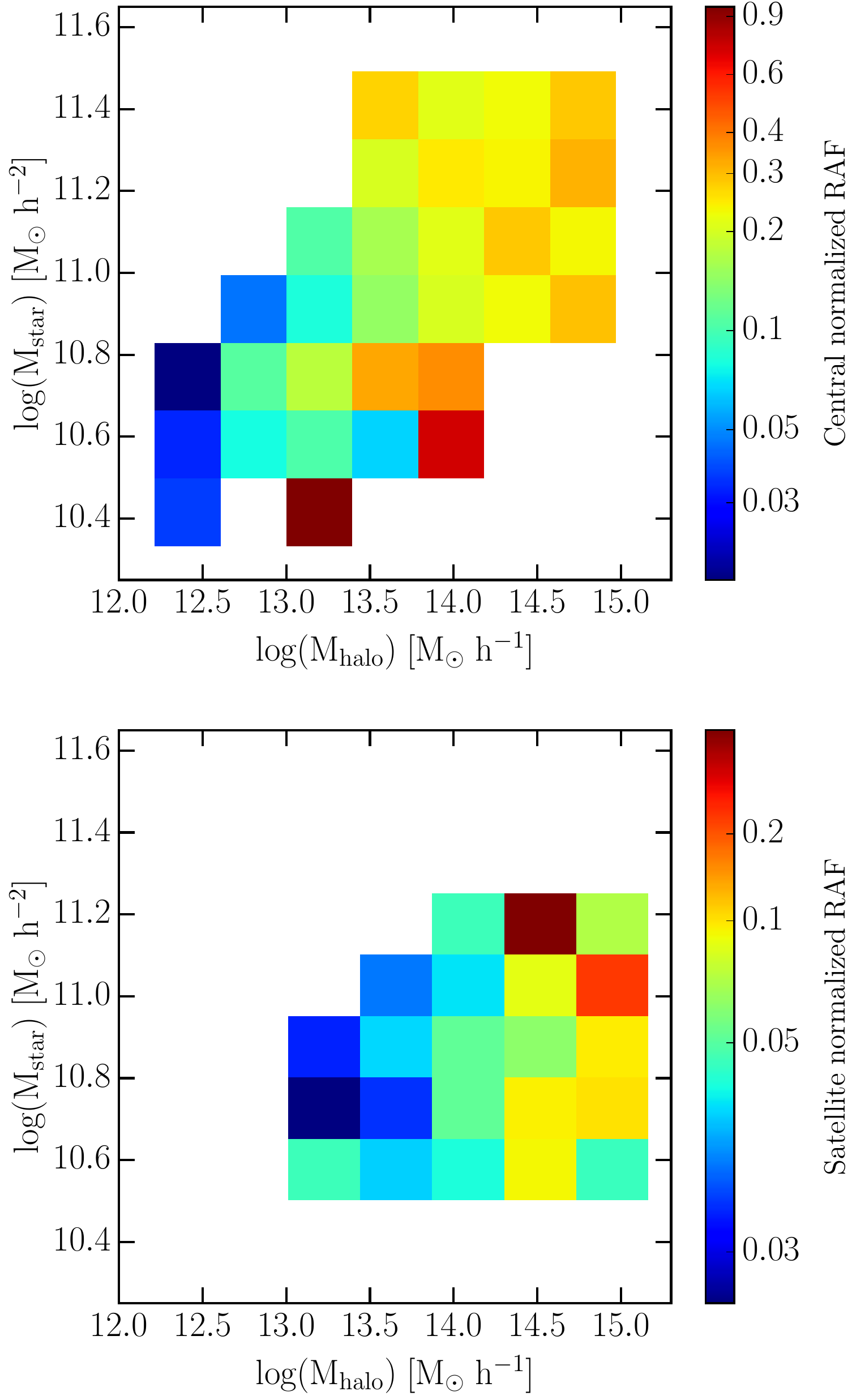}
\caption{
The dependence of normalized RAF on dark matter halo mass and stellar mass for central  (top panel) and satellite (lower panel) galaxies.  At fixed stellar mass, the normalized RAF is positively correlated with halo mass, for both central and satellite galaxies.  
While by definition, at fixed stellar mass, the normalized RAF sums up to unity,
it is  not meaningful to look for variation along stellar mass axis at fixed halo mass.
}
\label{fig:normraf}
\end{figure}

Fig.~\ref{fig:normraf} shows the normalized RAF for both central and satellite galaxies.  There is still a dependence on halo mass after the effect of the stellar mass--halo mass relation is removed.
Fitting a linear relation between the normalized RAF and logarithm of halo mass for each of the stellar mass bin, we find that all the slopes are positive.
To compute the uncertainties in the slope of the normalized RAF--halo mass fits, we generate 1000 bootstrap resampling and repeat the fitting process.  The halo mass dependence of the normalized RAF is highly significant ($>3\sigma$) for all but the most massive central galaxies (with the latter being at $2\sigma$ level, which is likely due to  small number statistics).
The above exercise is repeated for satellite galaxies.  Again we find the normalized RAF is positively correlated with halo mass at high significance.

We have thus established that the radio activity strongly depends on the halo mass the galaxies reside in.
This is consistent with the findings of some recent studies (e.g., \citealt[][]{mendez16,shen17}); for example,
\citet{ching17} find that LERGs are found in higher mass halos than a control sample of radio-quiescent  galaxies, indicating the importance of halo mass in triggering radio activity.
As the fueling of SMBHs takes place at a scale much smaller than that of a dark matter halo, we seek to find out what it is in massive halos that promotes the radio activity.  The possibilities include elevated rates of galaxy interaction, and the properties of ICM.

\begin{figure}
\hspace{5mm}
	\includegraphics[width=0.8\columnwidth]{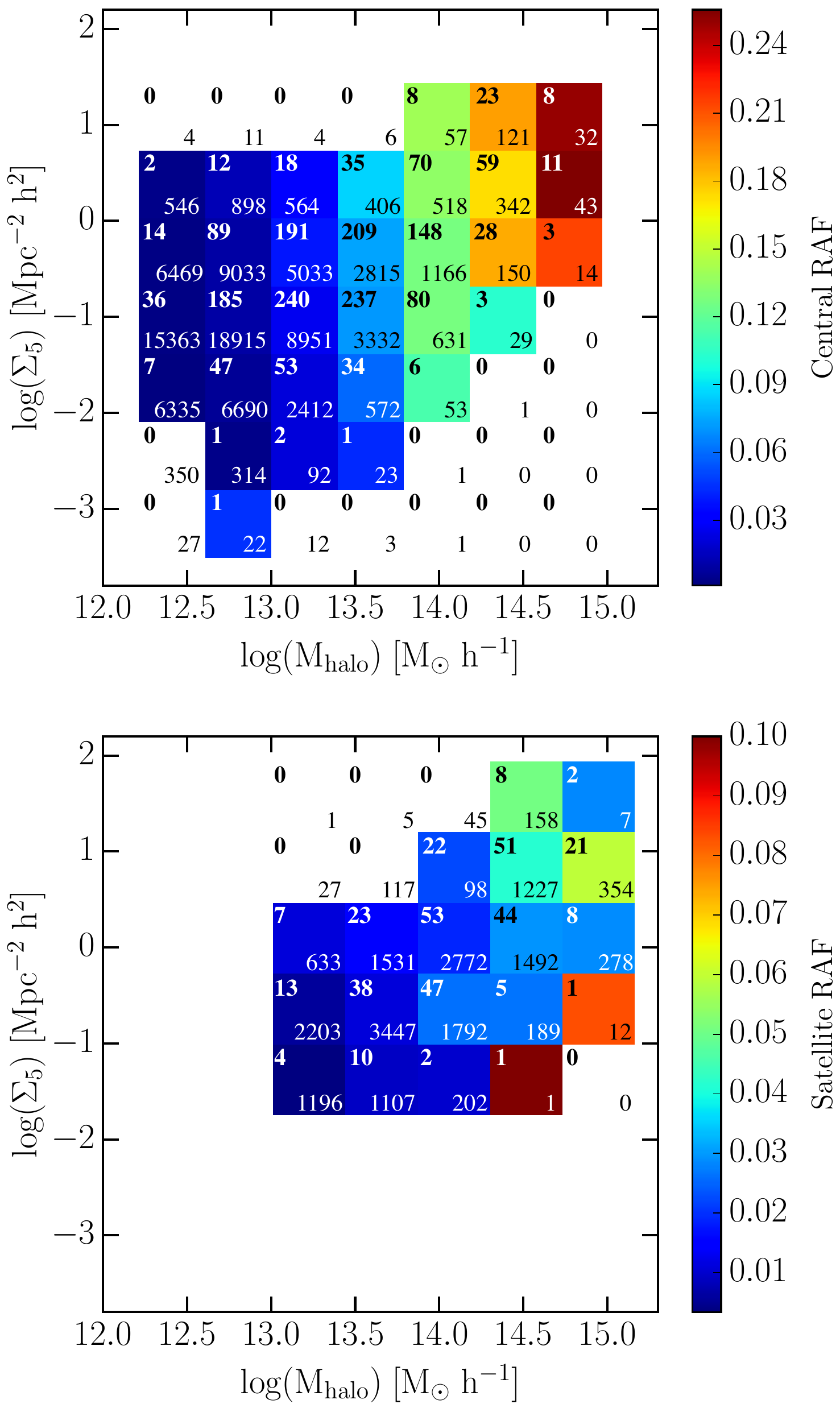}
    \caption{
The dependence of RAF on dark matter halo mass and local galaxy density, $\Sigma_5$, for central (top panel) and satellite (bottom panel) galaxies.  The RAF is largely independent of the local density.  We note that the galaxy sample used here is primarily composed of luminous galaxies (all in our main sample), and thus the local density measurements may not be representative, however (see Fig.~\ref{fig:mhsignorm}).
The meaning of the numbers in each of the ($M_h$, $\Sigma_5$) bins is the same as in Figure~\ref{fig:mhms}.
}
\label{fig:mhsig}
\end{figure}

\begin{figure}
	\includegraphics[width=0.8\columnwidth]{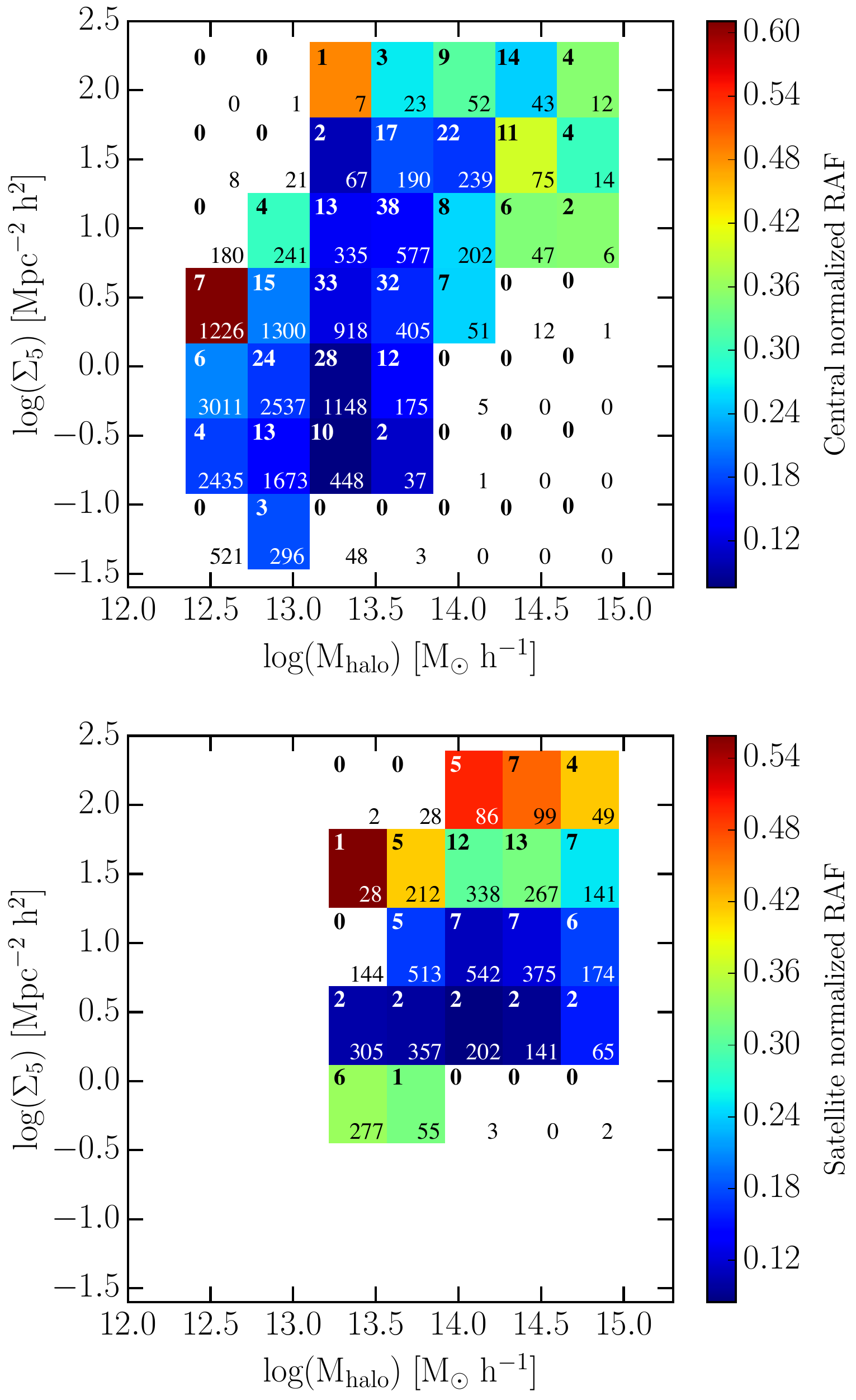}
    \caption{The dependence of normalized RAF on dark matter halo mass and local galaxy density for central (top panel) and satellite (bottom panel) galaxies.  Unlike in Fig.~\ref{fig:normraf}, here the RAF is normalized by fixing the halo mass.  The galaxy sample used here is volume-limited to $z=0.092$ (see Table~\ref{tab:sdssrg}), different from that used in Fig.~\ref{fig:mhsig}, as we would like to include lower luminosity galaxies in the calculation of $\Sigma_5$.  It can be seen that while the satellite RAF  depends on the local density, this does not seen to be the case for centrals.
The meaning of the numbers in each of the ($M_h$, $\Sigma_5$) bins is the same as in Figure~\ref{fig:mhms}.
}
\label{fig:mhsignorm}
\end{figure}

In Fig.~\ref{fig:mhsig} we show the RAF dependence in the local galaxy density $\Sigma_5$ $vs.$ halo mass plane.  Here $\Sigma_5$ is the surface density over an area containing the fifth nearest neighbor (also satisfying $M_r^{0.1}\le -21.57$ and velocity difference $\le 1000$\,km/s), using spectroscopic redshifts from SDSS;
it is a popular estimator of local density as it is related to the dark matter halo density \citep{sabater13}.
Our conclusions remain unchanged if we use the density constructed with third nearest neighbor, $\Sigma_3$.
There are hints of a weak dependence on the local galaxy density, although the dependence does not seem to be monotonic with respect to $\Sigma_5$.
We note that, however, the  selection of neighbors is limited to very luminous galaxies, which would bias us against effects of minor mergers or interactions with less massive companions.  We have thus repeated the exercise with a different volume limited sample (out to $z=0.092$,
with the same radio luminosity range, see Table~\ref{tab:sdssrg}), computed $\Sigma_5$ with less luminous ($M_r^{0.1}\le -20.27$) neighbors, 
and derived the normalized RAF [this time calculated at fixed halo mass:
$\tilde{f}(\Sigma_5|M_h)=f(\Sigma_5,M_h)/\int f(\Sigma_5,M_h) d\Sigma_5$],  
to remove the potential correlation between halo mass and local galaxy density.
The results are presented in  Figure~\ref{fig:mhsignorm}.  Compared to the trends shown in Figure~\ref{fig:mhsig}, for central galaxies in high mass halos ($M_h\ge 10^{14}h^{-1}\,M_\odot$), the effects of the local density on the RAF appears to be weakened.
The effect of the local density on satellites is more apparent, although we note the significance of the result is hampered by the small number of RGs, especially in densest regions (at $ \Sigma_5>100 h^{2}\,$Mpc$^{-2}$, we only have $4-7$ RGs in each cell).
For more quantitative assessment of the importance of the local galaxy density, we shall employ the LR analysis in the next section.

\subsection{Quantitative Analysis with Logistic Regression}
\label{sec:sdsslr}

We have shown that the RAF strongly depends on both $M_*$ and $M_h$,
with some hints of dependence on the local galaxy density.
Here we turn to LR as an independent way of evaluating the importance of these physical quantities on the radio activity of a given galaxy population.
We
simplify the radio activity into an ``on'' or ``off'' state (i.e., when the radio luminosity is within the adopted luminosity range or not), which allows us to transform the question into a classification problem, making the LR an appropriate analysis tool.
Equivalently, we are then asking: which of these physical properties are most predictive of the radio activity?

The goal of a regression analysis is to study the relationships between a response variable $Y$ and an array of $m\,$($\ge 1$) regressors $\bold{X} = (X_1, ... X_i, ... X_m)$
in a specified functional form $Y=f(\bold{X}) + \epsilon$. In our case, 
the response variable $Y$ is a binary, taking values of $Y=1$ for radio-loud and $Y=0$ for radio-quiescent.\footnote{Strictly speaking, as our RGs are selected to have $\log P_{1.4}=23.3-25.5$, more powerful RGs would have $Y=0$ and so this state should be understood as the complement of the ``low-power'' state.}
The regressors considered here are $M_{*}, M_{h}$ and $\Sigma_5$ (or $\Sigma_3$). We apply the standard LR model as the following:
\begin{equation}\label{eq:LR}
Pr(Y=1|\bold{X}) = f(\bold{X}) + \epsilon = \frac{e^{\beta_0+\beta_1 X_1 + \ldots+\beta_i X_i+ \ldots + \beta_m X_m }}{1+e^{\beta_0+\beta_1 X_1 + \ldots+\beta_i X_i+ \ldots + \beta_m X_m}} + \epsilon,
\end{equation}
where $Pr(Y=1|\bold{X})$ stands for the probability of a galaxy to be in the radio-loud state given its physical parameters $\bold{X}$, and $\epsilon$ represents random observational error.
Given the observational data of $n$ galaxies $(\bold{x_1}, y_1), (\bold{x_2}, y_2), ..., (\bold{x_n}, y_n)$ as input, we fit Eq.~\eqref{eq:LR} using the {\tt bestglm} package in {\it R} and derive the best fit model parameters $\hat{\beta_0}$, and $\hat{\bold{\beta}} = (\beta_1, \ldots,  \beta_m)$. 
A larger $|{\beta_i}|$ indicates a stronger effect of the regressor $X_i$ on $Y$.
It can be seen that if ${\beta_i}=0$, $X_i$ does not have any effect on the final value of $Pr(Y=1|\bold{X})$. 
Conventionally, the statistical significance level of regressor $X_i$'s effect on $Y$ is characterized via the $z$-value, defined as the ratio of estimated $\hat{\beta_i}$ to its standard error; a larger $|z|$ implies stronger evidence against the null hypothesis of $\beta_i=0$.
Under the assumption that ${\beta_i}$ is asymptotically Gaussian, there is a direct link between the $z$-value and $p$-value on the hypothesis test of whether ${\beta_i}=0$ or not. A $p$-value of 0.05 corresponds to a 95\% confidence interval for ${\beta_i}$ not overlapping with zero.

Aiming at selecting a model with predictors (regressors) that are truly associated with nuclear radio activity, we apply the {\it best subset selection} method to identify the best model. 
Generally, a model involving more regressors has more degrees of freedom to fit the data well, but may suffer from overfitting. On the other hand, a model with fewer regressors is more stable, but may lose its power to predict the behavior of $Y$ given all the available information.
The best subset selection method identifies the best model (among all 7 possible regressor combinations of $M_*, M_h,$ and $\Sigma$) as the one that minimizes the Bayesian Information Criterion ($BIC$), defined as
	\begin{equation}\label{eq:BIC}
	BIC \equiv \frac{1}{n} \left(RSS + {\rm log}(n)\, d\, \hat{\sigma}^2 \right),
	\end{equation}
where $RSS$ is the residual sum of squares, $d$ denotes the number of predictors used in the model, $n$ is the total number of galaxies, and $\hat{\sigma}^2$ is an estimate of the variance of observational error $\epsilon$ shown in Eq.~\eqref{eq:LR}.
The minimum of $BIC$ is reached by balancing the goodness of fit ($RSS$) and the degree of freedom $d$.

After the best model is identified, we can then evaluate the $z$- and $p$-values for each of the predictors in the best model. 
As mentioned above, the local density calculation for the $z\le 0.15$ sample is based on luminous galaxies, and thus may not be representative of true values.  We therefore focus on the $z\le 0.092$ sub-sample here.
For central galaxies, the best model consists of $M_h$ and $M_*$, irrespective of the local density definition ($\Sigma_5$ or $\Sigma_3$).  
As for satellite galaxies, LR prefers a model consisting of $M_*$ and local density (either of $\Sigma_5$ and $\Sigma_3$).
The resulting $z$- and $p$-values,  together with the best-fit $\beta$ parameter, of these models are shown in Table~\ref{tab:lr_sdss} (for completeness, we present results for both the main sample and the $z\le 0.092$ sub-sample). 
These findings are consistent with the qualitative trends revealed by the RAF plots (Figures~\ref{fig:mhms}--\ref{fig:mhsignorm}).

For the $z\le 0.15$ sample, we see from Table~\ref{tab:lr_sdss} that the $\beta$
values associated with $M_h$ are significantly different for the centrals and
satellites.  The same also applies to that associated with $M_*$.  Therefore we can 
conclude that the distribution of RAF for centrals and satellites in the $M_h$ {\it vs.} $M_*$ plane must be different.

\begin{table}
    \centering
    \caption{Best models based on the LR analysis}
    \label{tab:lr_sdss}
    \begin{tabular}{rcrr}
        \hline
 \multicolumn{2}{c}{$z\le 0.15$, central galaxies} & \\
 \hline
 Parameter   & $\beta$                         &    $z$-value     &    $p$-value     \\ \hline
 $M_h$        &        1.411$\pm$0.065     &    $21.72$        &    $<2\times 10^{-16}$ \\
 $M_*$         &       3.406$\pm$0.171     &    $19.97$        &    $<2\times 10^{-16}$     \\    \hline
 \multicolumn{2}{c}{$z\le 0.15$, satellite galaxies} & \\
 \hline
 Parameter   & $\beta$                    &   $z$-value    &    $p$-value     \\ \hline
 $M_h$        &   0.876$\pm$0.125  &   $7.03$        &    $<2\times 10^{-12}$ \\
 $M_*$         &  4.736$\pm$0.330   &   $14.36$      &    $<2\times 10^{-16}$     \\
 \hline \hline
 \multicolumn{2}{c}{$z\le 0.092$, central galaxies} & \\
 \hline
 Parameter   & $\beta$                         &    $z$-value    &    $p$-value     \\ \hline
 $M_h$         &  1.819$\pm$0.110    &    $16.57$       &    $<2\times 10^{-12}$ \\
 $M_*$         &   1.925$\pm$0.255   &     $7.57$        &    $<4\times 10^{-14}$     \\    \hline
 \multicolumn{2}{c}{$z\le 0.092$, satellite galaxies} & \\
 \hline
 Parameter     & $\beta$                         &  $z$-value    &    $p$-value     \\ \hline
 $M_*$           & 2.814$\pm$0.475  & $5.92$        &    $<3\times 10^{-9}$ \\
 $\Sigma_5$  & 0.958$\pm$0.173   & $5.55$        &    $<3\times 10^{-8}$     \\    \hline
    \end{tabular}
 \end{table}

The importance of interaction in triggering radio activity has been noted in a few studies \citep[e.g.,][]{sabater13,pace14}.  
Our distinction of central and satellite galaxies has enabled us to attribute this environmental factor (and the implied enhancement of tidal interactions) particularly to radio AGN phenomenon in satellites.
Using a sample of galaxies in spectroscopically confirmed pairs, \citet{ellison15} show that once halo mass and stellar age of galaxies are controlled, major mergers do not enhance the (low-excitation) radio activity,  and thus make the conjecture that minor mergers or/and accretion from the surrounding medium could be possible external gas fueling mechanisms (see also \citealt{karouzos14}).
Given that the majority of our satellite RGs live in cluster-scale halos (75\% are in halos with $M_h\ge 10^{14} M_\odot$), and that the high velocity dispersion in clusters makes both merger rates and gas accretion rates low for satellites, however,
it seems that tidal interactions are the most probable channel for bringing external gas into nuclear regions of satellites.

Although it is possible that enhanced interaction among galaxies in dense regions may partly contribute to the triggering of radio activity in satellite galaxies, we note that the majority ($\sim 85\%$) of RGs are central galaxies, and that the range of the local density of central galaxies is similar to that of satellites,
therefore in massive halos there must be other, presumably major, environmental factors  that promote nuclear radio activity -- the most obvious contender omitted in our analysis so far is the hot diffuse ICM.
In the next Section we thus explore in detail the dependences of radio activity on local ICM properties.

\section{Local ICM Properties} 
\label{sec:icm}

To investigate the effect of the ICM on triggering of the radio activity, we make use of the
X-ray measurements in the ACCEPT 
(Archive of Chandra Cluster Entropy Profile Tables) 
database \citep{cavagnolo09}, which is an attempt to homogeneously analyze {\it Chandra} observations of about 230 galaxy clusters. 
For each cluster, the database provides, as a function of distance from the cluster center, electron density $n_e$, pressure $p$, temperature $T_X$, entropy  (defined as $K\equiv T_X n_e^{-2/3}$), cooling time $t_{\rm c}$, and enclosed gravitational mass, from which we infer the free-fall time $t_{\rm ff} = \sqrt{2r^3/GM(<r)}$.  In addition, the database also provides the location of the X-ray emission peak (which we take as the cluster center), and the global mean X-ray temperature $\overline{T}$. 
By assuming azimuthal symmetry, we thus know the local ICM properties of every member galaxy given its distance from the cluster center.

In the redshift range $z=0.03-0.32$, there are 54 ACCEPT clusters within the final SDSS imaging footprint (i.e., DR8) with  ICM measurements that allow for a robust determination of $t_{\rm ff}$\footnote{In practice, we keep only clusters with monotonically varying pressure and $t_{\rm ff}$ profiles.}, which will be referred to as the ACCEPT-SDSS subsample.  We can thus use the SDSS (imaging and spectroscopic) data to identify cluster members and study the correlations between the radio activity and the local ICM properties.  
We have visually inspected these clusters to identify the BCGs.
As for the member galaxies, depending on the redshifts of the clusters, the treatments are somewhat different.
For clusters at $z\ge 0.1$, we make use of the membership probability $P_{\rm mem}$ as given by the redMaPPer algorithm \citep[][note that all ACCEPT-SDSS clusters have a counterpart in the redMaPPer cluster sample]{rykoff14}.  We regard a galaxy to be a potential cluster member if its $P_{\rm mem}\ge 0.8$. 
As the redMaPPer cluster sample is incomplete below $z=0.1$, for the ACCEPT-SDSS clusters at $z<0.1$, we make use of the spectroscopic redshifts and galaxy color to identify members.  Specifically, we regard as members those galaxies within 3000 km/s from the cluster restframe, or those with a restframe $g-r$ color consistent with the red sequence and are within a projected distance of 1.2 Mpc from the cluster center.  We only consider galaxies more luminous than $M_r^{0.1}=-21.27$, which is about the characteristic magnitude of the galaxy luminosity function \citep{blanton03b}.
Absolute magnitudes, restframe colors, and stellar masses are again calculated using {\tt kcorrect}.
Our results below do not depend sensitively on our choice of parameters for selecting cluster members.
Finally, we remove a few extreme outliers with nonsensical measurements in physical properties such as stellar mass, pressure, etc.

We have run our RG finding algorithm on the ACCEPT-SDSS clusters, 
with a flux limit of $1\,$mJy that is suitable for FIRST, 
and have visually inspected all potential optical-radio matches to finalize our cluster RG sample (see Table~\ref{tab:accept}).
As radio AGN are predominantly hosted by red galaxies  (for example, about $86\%$ of the RGs with $\log P_{1.4} \le 25.5$ in the L10 sample lie on the red sequence; see also \citealt{janssen12}), our reliance on the redMaPPer membership assignment, which only considers red galaxies, should not bias ourselves against RGs.
In total, there are 509 satellite galaxies that lie within the coverage of {\it Chandra} observations, out of which 24 have radio power $\log P_{1.4} \ge 23.5$, a limit chosen to ensure our sample is volume limited.  Of the 54 BCGs, 16 are RGs.  This fraction is consistent with that found by \citet{lin07}.

Among the available physical properties associated with each galaxy, we consider a total of six  parameters here: cluster mass (using the global mean X-ray temperature as a proxy), stellar mass, ICM pressure, entropy, cooling time $t_{\rm c}$, and the ratio of cooling time over free-fall time, $t_{\rm c}/t_{\rm ff}$. One physical attribute unfortunately omitted here is the local galaxy density, as we do not possess sufficient number of spectroscopic redshifts for the ACCEPT clusters. 
Given the small size of the ACCEPT-SDSS RG sample,
and that many of the ICM-related parameters are highly correlated, we are not able to perform the best subset selection to pick up the most important predictors, as has been done in Section~\ref{sec:sdsslr}, because in the present case any potential statistical fluctuations could change the rank of {\it BIC} values among different models. 
Instead, our strategy is to first reduce the dimensionality of the parameter space by the application of the
 the {\it principle component analysis} (PCA) among the six parameters, and investigate the effects of the first few dominant principle components (PCs) on the nuclear radio activity.  We further strengthen the findings with the LR analysis.
Below we first present results for the BCGs, then show those for the satellites.

\subsection{BCGs}

In finding out the PCs , it is necessary to scale each of our physical parameters $P$ 
to $Q \equiv \frac{P - \bar{P}}{\sigma_{P}}$, where $\bar{P}$ and $\sigma_{P}$ are the mean and the standard deviation of the parameter. 
Such a normalization ensures that the derived PCs are not dominated by a single variable that has the largest standard deviation over the parameter space.

\begin{figure*}
	\includegraphics[width=2\columnwidth]{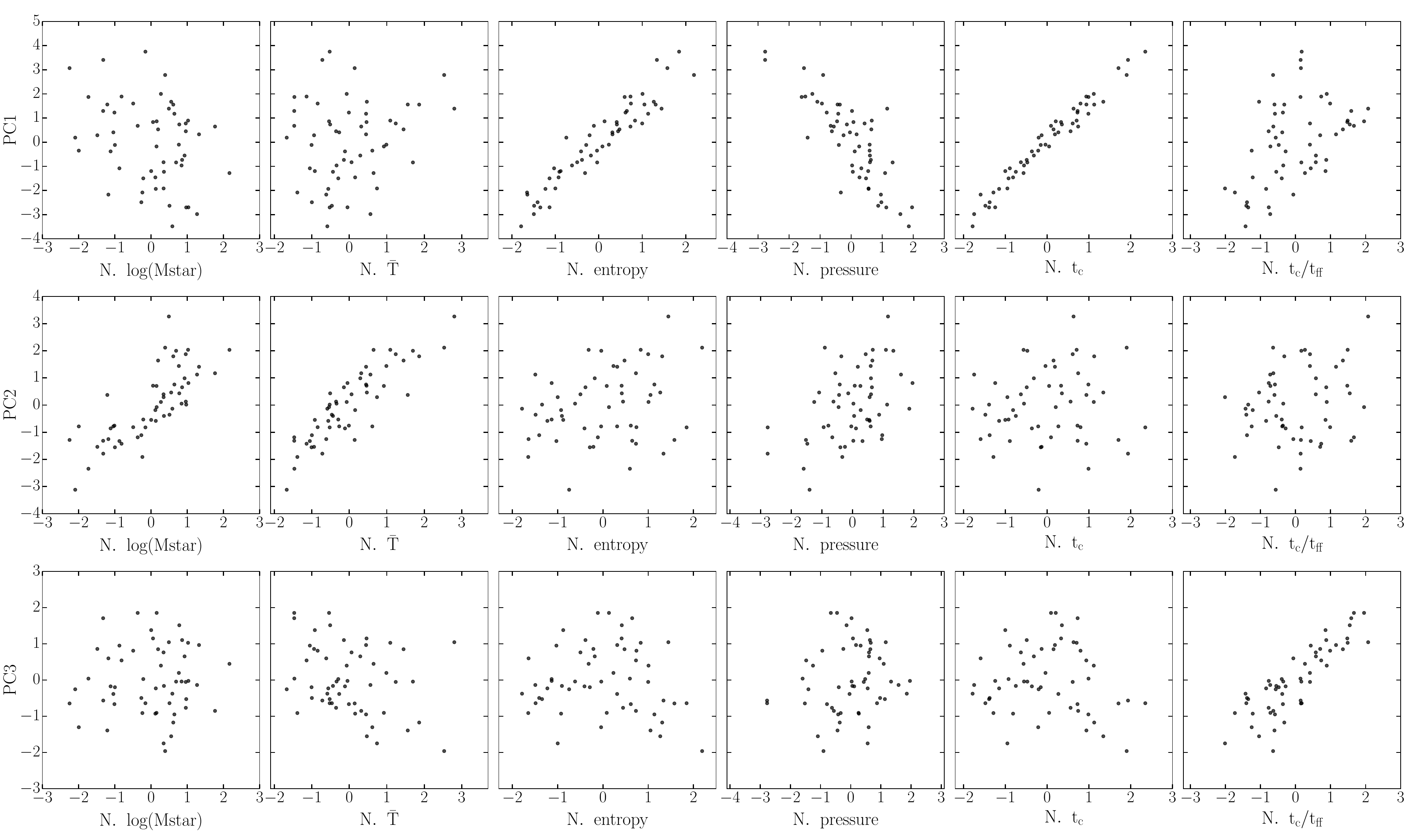}
    \caption{Relation between the parameters and the three dominant principal components for BCGs. The letter ``N'' in the abscissa stands for ``normalized''.
}
\label{fig:PCA_BCG}
\end{figure*}

For BCGs, the PCA indicates that only
three PCs are needed to explain up to 90\% of variance across the six dimensional parameter space. 
As shown in Table~\ref{tab:PCA_BCG}, 
the first PC (hereafter PC1) accounts for $\sim48\%$ of the variance, with its effect mostly on the variation of entropy, pressure and $t_{\rm c}$. 
More specifically, we have 
\begin{equation}\label{eq:PC1_BCG}
\begin{aligned}
\begin{split}
PC1 = & 
-0.18\ \left( \frac{\tilde{M_*}-11.33}{0.30} \right)		
+0.10\ \left( \frac{\overline{T}-6.03}{2.40} \right)				\\
&+0.56\ \left( \frac{\tilde{K}-1.96}{0.50} \right)
-0.48\ \left( \frac{\tilde{p}+9.92}{0.54} \right)			\\
&+0.58\ \left( \frac{\widetilde{t_{\rm c}}-0.48}{0.63} \right)
+0.27\ \left( \frac{{t_{\rm c}/t_{\rm ff}}-2.01}{0.42} \right),		
\end{split}
\end{aligned}
\end{equation}
where a tilde denotes a quantity in logarithm.
A higher entropy, lower pressure, and larger $t_{\rm c}$ lead to a larger PC1, as revealed in the top row of Fig.~\ref{fig:PCA_BCG}.
The second PC (PC2) largely depends on the global temperature $\overline{T}$ and stellar mass, as shown in the middle row of Fig.~\ref{fig:PCA_BCG};
the third PC (PC3) is mainly driven by $t_{\rm c}/t_{\rm ff}$ (see the bottom row of Fig.~\ref{fig:PCA_BCG}).

\begin{table*}
	\centering
	\caption{PCA results for 54 BCGs in the ACCEPT-SDSS sample}
	\label{tab:PCA_BCG}
	\begin{tabular}{lccccccc} 
		\hline
		 &  variance	& $\log M_{\rm star}$	&	$\overline{T}$	&	$\log K$		& $\log p$ 	& $\log t_{\rm c}$ 	& $ t_{\rm c}/t_{\rm ff}$ \\
		 & accounted & & & & & & \\
		\hline
		PC1	&         48\%   		&        	$-0.18$			&    	0.10			& 		0.56		& 	$-0.48$		& 	0.58			& 	0.27\\
                 PC2	&         28\%   		&     		0.60			&	0.67			&		0.22		&	0.35		& 	0.07			&	0.16					\\ 

                 PC3	&        14\%   		&       	0.05			&	$-0.31$			&		$-0.09$		&		0.18		&	$-0.12$		&	0.91					\\ 
		\hline
	\end{tabular}
\end{table*}

\begin{figure*}
	\includegraphics[width=1.7\columnwidth]{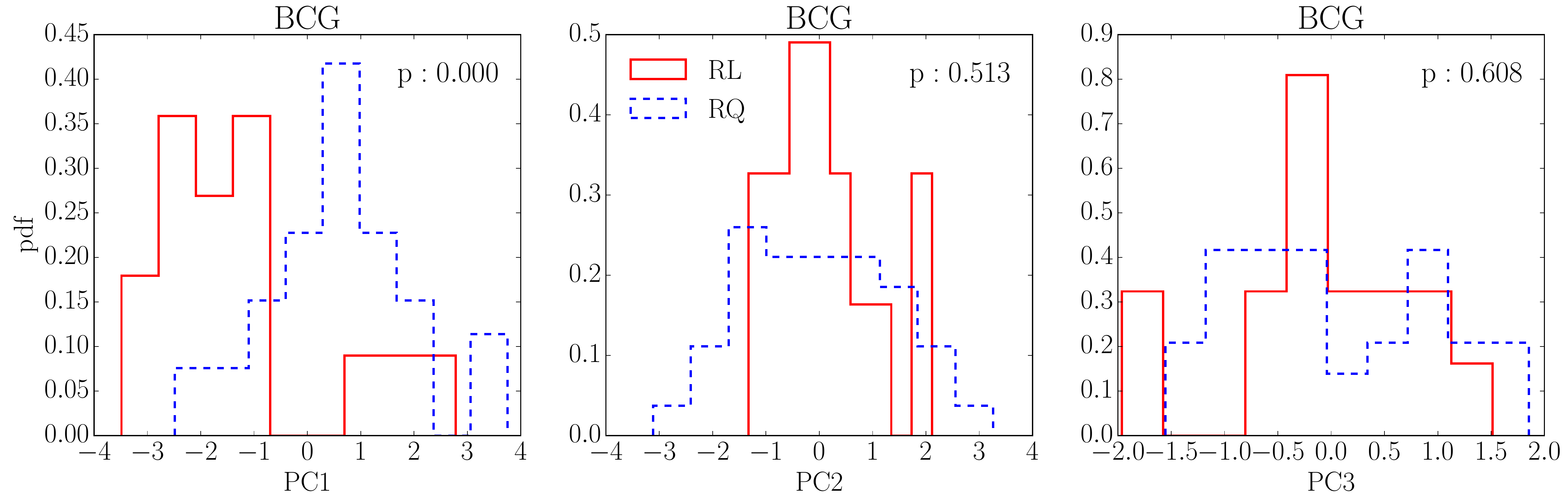}
    \caption{The probability density distribution of PC1, PC2, and PC3 for radio-loud (red) and radio-quiescent (blue) BCGs.  The $p$-values obtained from a KS test comparing the two distributions are also shown in each panel.  For PC1, the low $p$ value indicates that the two distributions are highly inconsistent with each other.}
\label{fig:cen_PCdist}
\end{figure*}

We can further examine the effects of the PCs on the radio activity of BCGs.
Fig.~\ref{fig:cen_PCdist} shows the probability density function (pdf) of each PC for radio-loud (red solid histograms) and radio-quiescent (blue dash histograms) BCGs. 
The $p$-values obtained from a Kolmogorov-Smirnov (KS) test comparing the two distributions are also shown in each panel.
It is clear that PC1 has the strongest effect, 
implying that an environment with low entropy, high pressure, and short cooling time is favorable  for triggering the nuclear radio activities.
For the distribution of PC2 and PC3, the large $p$-value reveals that there is no significant difference between RGs and normal galaxies.  But here we note that,
the relatively weak effect of PC2 on the BCG radio activity may be due to the limited mass range our clusters span.
As we have seen from Section~\ref{sec:halomass}, both stellar mass and halo mass (basically the PC2 here) are related to the RAF of BCGs.

We have also performed the LR analysis on the ACCEPT-SDSS BCGs.  Given the small sample size, 
here we only consider the one-predictor LR model [i.e.~the case when $\bold{X}$ in Eq.~\eqref{eq:LR}  is a scalar] for each of the regressors,
 and tabulate the resulting $z$- and $p$-values in Table~\ref{tab:LR_BCG}.
The results largely agree with that given by the PCA, that is, entropy, pressure, and cooling time all show stronger effects on the RG activity than other covariates.
There is some hint of cooling time and entropy being more important than the pressure.

\begin{table}
	\centering
	\caption{LR statistics for BCGs in the ACCEPT-SDSS sample}
	\label{tab:LR_BCG}
	\begin{tabular}{lrr} 
		\hline
Parameter				&	$z$-value	&	$p$-value	\\  \hline
$\log M_{\rm star}$		&	1.89		&	0.059	\\ 
$\overline{T}$			&	$-0.54$		&	0.592	\\ 
$\log K$			&	$-3.16$		&	0.002	\\ 
$\log p$			&	2.61		&	0.009	\\ 
$\log t_{\rm c}$			&	$-3.18$		&	0.001	\\ 
$t_{\rm c}/t_{\rm ff}$	&	$-2.08$		&	0.038	\\
		\hline
	\end{tabular}
\end{table}

Admittedly a sample size of only 54 limits the statistical significance of our inference on the predicting power of physical properties on the radio activity.
For  BCGs, we could increase the sample size  by loosening the requirement that the clusters should lie within the SDSS footprint; rather, we only need to be able to identify the BCGs robustly, and to be able to measure the radio properties of the BCGs.
To this end,
we expand our cluster sample to include all  ACCEPT clusters lying at declination $\delta>-40$\,deg and at $z<0.2$, and use Wide-field Infrared Survey Explorer \citep[WISE;][]{wright10} and Two-Micron All-Sky Survey \citep[2MASS;][]{skrutskie06} data to identify the BCGs, for clusters lying outside of the SDSS footprint.  The WISE channel 1 luminosity is used as a proxy of stellar mass (e.g., \citealt{lin13}).
Data from NVSS is used to select radio-loud BCGs and measure their fluxes.
For the 105 central galaxies in this ACCEPT-WISE sample, 
both the PCA and LR results are largely similar to that of ACCEPT-SDSS BCGs.  Our conclusions are therefore robust against the limited sample size.

\subsection{Satellites}

\begin{figure*}
	\includegraphics[width=2\columnwidth]{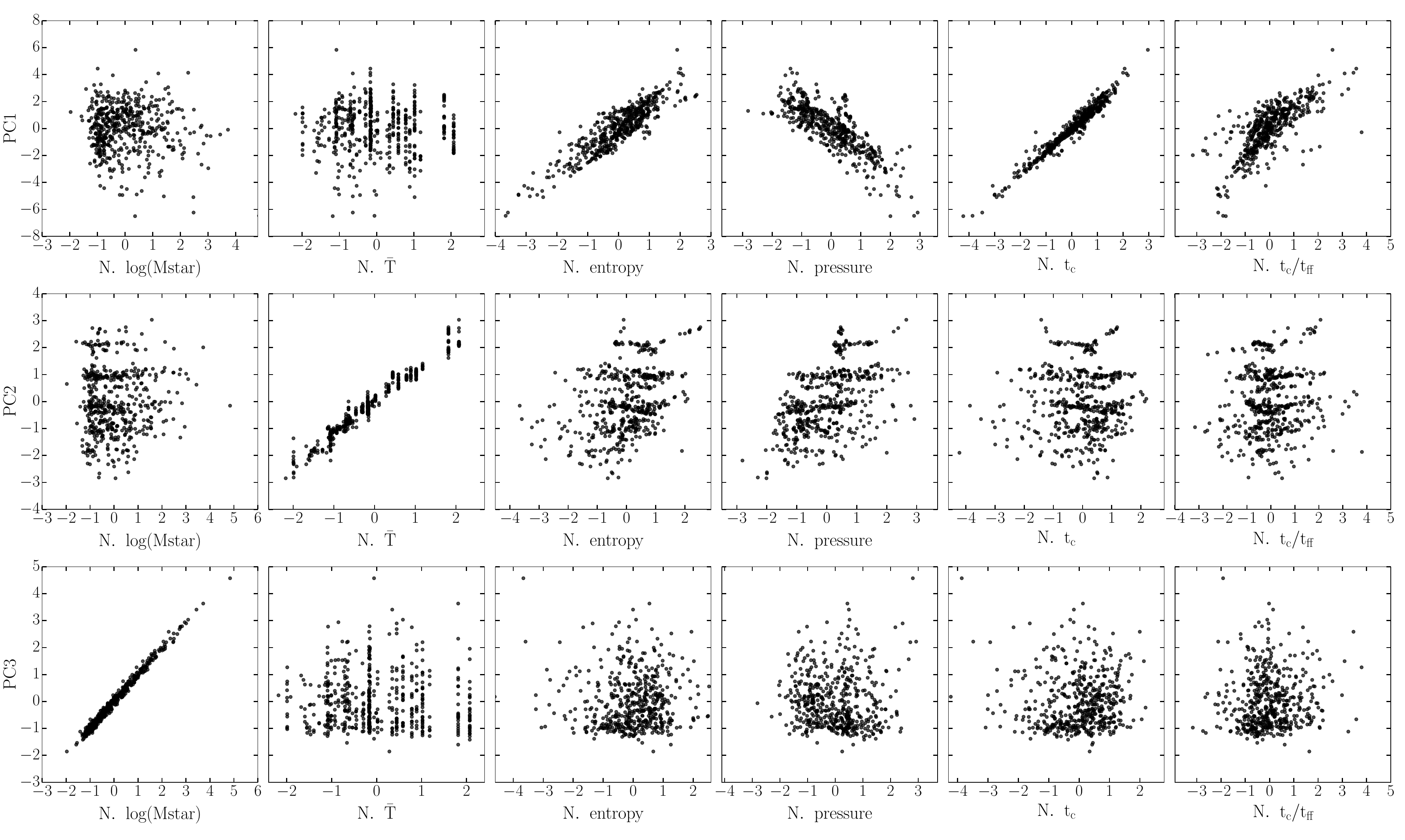}
    \caption{Same as Figure~\ref{fig:PCA_BCG}, but for satellites.
}
\label{fig:PCA_SAT}
\end{figure*}

The PCA results for satellites in the ACCEPT-SDSS sample are summarized in Table~\ref{tab:PCA_SAT}. 
As is the case for the BCGs, the first three PCs capture $\sim 90\%$ of variance over the six dimensional parameter space.
However, there are slight differences in the physical information as revealed by the PCs of satellites. 
As shown in Fig.~\ref{fig:PCA_SAT}, besides entropy, pressure, and cooling time, the PC1 of satellites has larger contribution from $t_{\rm cool}/t_{\rm ff}$ compared with that of BCGs. The PC2 of satellites shows mainly the effect of  $\overline{T}$, and PC3 is totally dominated by stellar mass. 
It is seen that the effect of stellar mass is quite decoupled from other (extrinsic) properties considered.

\begin{table*}
	\centering
	\caption{PCA results for 509 satellites in the ACCEPT-SDSS sample}
	\label{tab:PCA_SAT}
	\begin{tabular}{lccccccc} 
		\hline
		 &  variance	& $\log M_{\rm star}$	&	$\overline{T}$	&	$\log K$		& $\log p$ 	& $\log t_{\rm c}$ 	& $ t_{\rm c}/t_{\rm ff}$ \\
		 & accounted & & & & & & \\
		\hline
PC1	&         50\%   		&        	$-0.05$			&    	$-0.02$			& 		0.52		& 	$-0.47$		& 	0.57			& 	0.43 					\\   

PC2	&         23\%   		&     		0.03			&	0.83			&		0.33		&	0.43		& 	0.03			&	0.09					\\ 

PC3	&        17\%   		&       	1.00			&	$-0.03$			&		$-0.01$		&	$-0.03$		&	0.01			&	0.07					\\ 
		\hline
	\end{tabular}
\end{table*}

\begin{figure*}
	\includegraphics[width=1.7\columnwidth]{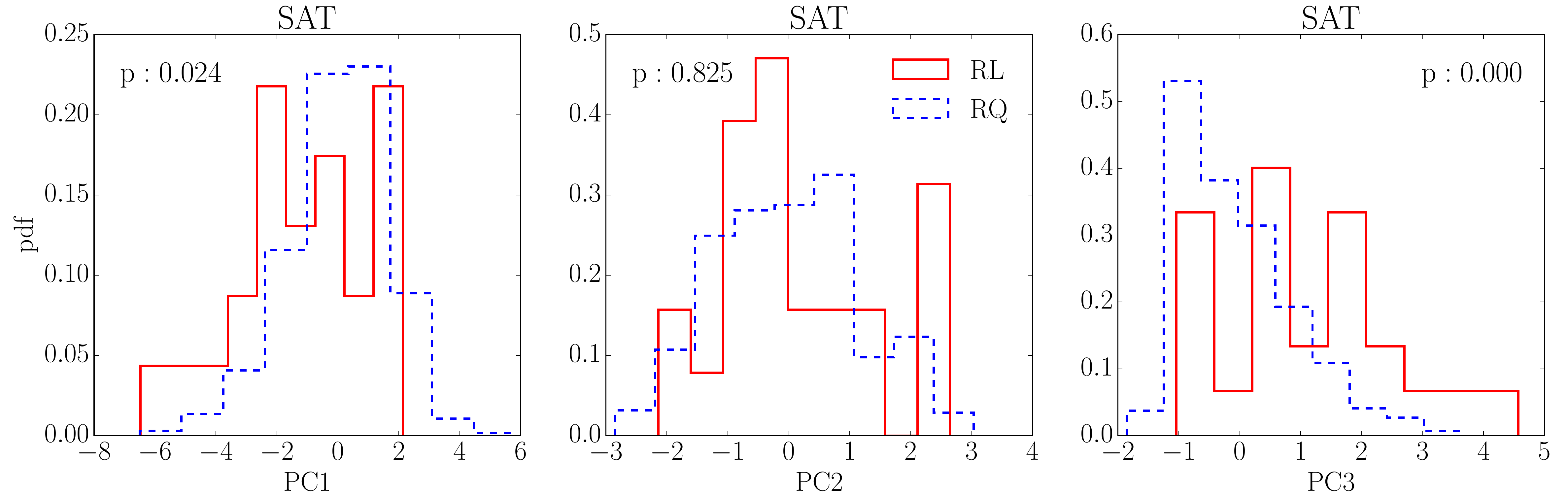}
    \caption{The probability density distribution of PC1, PC2, and PC3 for radio-loud (red) and radio-quiescent (blue) satellites.  The $p$-values obtained from a KS test comparing the two distributions are also shown in each panel.  For PC3, the low $p$ value indicates that the two distributions are highly inconsistent with each other.}
\label{fig:sat_PCdist}
\end{figure*}

Fig.~\ref{fig:sat_PCdist} show the pdf of PCs for radio-loud (red solid histograms) and radio-quiescent (blue dash histograms) satellites. 
It is apparent that the radio-loud and radio-quiescent satellites show different distributions in PC1 and PC3 (with very low $p$ values), with active satellites showing preference to lower PC1  (low entropy, cooling time, and $t_{\rm cool}/t_{\rm ff}$, and high pressure) and higher PC3 (higher stellar mass) values. This  is consistent with the findings in Section~\ref{sec:halomass}.

In Table~\ref{tab:LR_SAT} we summarize the significance for each of the parameters via the LR analysis for the satellites. 
We see that stellar mass is the most powerful predictor for RG activity, although cooling time, entropy, and pressure all play important roles. 
It should be kept in mind that the number of RGs used in our satellite analysis is rather small ($\lesssim 5\%$ of the satellites are radio-loud); a larger cluster sample will be needed to firmly establish these conclusions.

\begin{table}
	\centering
	\caption{LR statistics for satellites in the ACCEPT-SDSS sample}
	\label{tab:LR_SAT}
	\begin{tabular}{lrr} 
		\hline
Parameter				&	$z$-value	&	$p$-value	\\ \hline 
$\log M_{\rm star}$		&	$5.07$		&	$<0.001$	\\ 
$\overline{T}$			&	$0.50$		&	$0.620$	\\ 
$\log K$			        &	$-2.91$		&	$0.004$	\\ 
$\log p$				&	$2.58$		&	$0.010$	\\ 
$\log t_{\rm c}$			&	$-3.16$		&	$0.002$	\\ 
$t_{\rm c}/t_{\rm ff}$		&	$-1.66$		&	$0.097$	\\	\hline
	\end{tabular}
\end{table}

\section{Summary and Discussion} 
\label{sec:disc}

We have outlined an analysis framework that combines a group and cluster sample that spans a wide range in halo mass, and a cluster sample that offers detailed ICM measurement.  Together with standard statistical analysis tools, the joint samples allow us to investigate the likely sources of nuclear activity in massive galaxies, making it possible to sort out the relative contributions from halo mass, stellar mass, and other physical properties such as local galaxy density, ICM entropy, cooling time.

In the first part of our analysis (Section~\ref{sec:halomass}), we have used a large sample of galactic systems to show that, for triggering the radio activity, stellar mass is an important factor, irrespective of 
the type of  galaxies (central v.s.~satellite).
The central galaxy RAF  additionally strongly depends on the mass of their host dark matter halo (such dependence is also present for the satellites but is weaker).  
As we do not find convincing evidence linking the elevated RAF in massive halos to the higher galaxy density therein (which in turn should correlate with the interaction rates) for central galaxies, the most likely culprit is the presence of hot gas in massive halos.
Thus in the second part of our analysis (Section~\ref{sec:icm}), using a cluster sample that provides spatially resolved measurements of ICM properties, we seek for the best predictor of radio activity among internal and environmental factors that could play a role in the evolution of a galaxy.

According to our results, entropy, cooling time, and pressure play important roles in the triggering of radio activity in galaxies residing in massive halos.
As two of these properties are intimately linked to gas cooling out of the hot ICM, a picture that emerges from our study is the following: 
whether or not a galaxy is a central, the more massive it is, the more likely it will be active in the radio.  The likely source of fuel for the SMBH is from stellar mass loss from evolved stars (e.g., AGB stars).  If the galaxy happens to be central in a massive halo where appreciable amount of ICM is present, then an extra fuel supply, likely gas cooling out of the hot ICM, could trigger more radio activity.  Confining pressure of the ICM further provides a ``working surface'' for the jets.

Although this picture is certainly not  new \citep[e.g.,][]{ciotti01,best05b,ho08,heckman14}, our analysis framework provides a (nearly) self-contained way to demonstrate it.  Furthermore, we do not find the ratio $t_c/t_{\rm ff}$ to be a critical parameter for the radio activity (particularly for BCGs), which seems to be at odds to the predictions from the precipitation model advocated by \citet{voit15}.

For our analysis, there are several aspects that can be improved or extended.  For example, for the ACCEPT cluster sample, we do not have a good handle on the local galaxy density.  We could have used clustercentric distance as a very crude proxy for the local density, which is obviously too simplistic.  Either spectroscopic data from surveys such as GAMA \citep[Galaxy And Mass Assembly;][]{driver11}, or good photometric redshifts with adequate background correction methods, or both, are needed to estimate the local density in a statistical fashion.
With a larger cluster sample, such as that from ACCEPT2, one could also consider high and low excitation radio galaxies separately, and gain more insight into the potentially different triggering mechanisms for these two types of AGN (e.g., \citealt{hardcastle07}).
In addition, applying our approach to dense spectroscopic surveys (e.g., GAMA), we could also study the triggering of AGNs selected by (optical) emission lines (e.g., \citealt{kauffmann03,sabater15}).
Finally, by combining data from surveys such as 
GAMA (or other dense-sampling spectroscopic surveys such as PRIMUS, Dark Energy Spectroscopic Instrument  and Prime Focus Spectrograph),  ACCEPT2, and Pan-STARRS \citep{chambers16} (or deeper imaging surveys such as Kilo Degree Survey, Dark Energy Survey, or Hyper Suprime-Cam), it may be possible to extend our analysis to intermediate redshifts ($z\sim 0.5-1$).  
Given that the cool-core properties of clusters, such as central density, entropy, and cooling time, have remained unchanged since $z\sim 1$ \citep{mcdonald13,mcdonald17}, we expect the RAF of BCGs to remain similar over cosmic time.  On the other hand, both the galactic interaction rates and the cold gas content in galaxies are expected to increase towards high-$z$, implying an elevated RAF among satellites.  It would thus be exciting to examine the triggering of radio AGN at different cosmic epochs (e.g., \citealt{williams15,lin17}).

One important omission in the current study is the spin of the SMBH.  Given the difficulty of estimating the magnitude of spin, it does not seem feasible to incorporate it in statistical analyses like ours, however.  Its relevance compared to stellar mass, or extrinsic properties such as entropy and cooling time, has to be assessed with more focused studies (e.g., \citealt{schulze17}), and is beyond the scope of the current paper.

\section*{Acknowledgments}

We thank an anonymous referee for helpful comments that have improved the clarity of the paper.
YTL acknowledges support from the Ministry of Science and Technology grants MOST 104-2112-M-001-047 and MOST 105-2112-M-001-028-MY3, and an Academia Sinica Career Development Award (2017-2021).  
This project received financial support from the Conselho Nacional de Desenvolvimento Cient\'{i}fico e Tecnol\'{o}gico (CNPq) through grant 400738/2014-7.
YTL thanks Mark Voit, Jenny Greene, Luis Ho, Ming Sun, Andrey Kravtsov, Roderik Overzier and Kai-Feng Chen for useful comments, and IH and LY for constant encouragement and inspiration.

Funding for the SDSS and SDSS-II has been provided by the Alfred P. Sloan Foundation, the Participating Institutions, the National Science Foundation, the U.S. Department of Energy, the National Aeronautics and Space Administration, the Japanese Monbukagakusho, the Max Planck Society, and the Higher Education Funding Council for England. The SDSS Web Site is http://www.sdss.org/.

The SDSS is managed by the Astrophysical Research Consortium for the Participating Institutions. The Participating Institutions are the American Museum of Natural History, Astrophysical Institute Potsdam, University of Basel, University of Cambridge, Case Western Reserve University, University of Chicago, Drexel University, Fermilab, the Institute for Advanced Study, the Japan Participation Group, Johns Hopkins University, the Joint Institute for Nuclear Astrophysics, the Kavli Institute for Particle Astrophysics and Cosmology, the Korean Scientist Group, the Chinese Academy of Sciences (LAMOST), Los Alamos National Laboratory, the Max-Planck-Institute for Astronomy (MPIA), the Max-Planck-Institute for Astrophysics (MPA), New Mexico State University, Ohio State University, University of Pittsburgh, University of Portsmouth, Princeton University, the United States Naval Observatory, and the University of Washington.


\appendix

\section{Radio Galaxy Samples}

Here we present the radio galaxy samples used in this work.  Table~\ref{tab:L10tab} is the parent sample for the RGs used in Section~\ref{sec:halomass}, which is obtained by combining the L10 RG catalog and the part of the \citet{best12} catalog in the area unique to SDSS DR7 (with respect to DR6).
The sample is complete to $M_r^{0.1}\le -21.27$, with radio flux $f\ge 3\,$mJy and at $z\le 0.3$.
In Table~\ref{tab:L10tab} we list the coordinates, redshift, 1.4 GHz flux and radio luminosity (assuming a spectral index of $-0.8$),  restframe magnitude in $r^{0.1}$ band, stellar mass, indication of whether the source is powered by an AGN, a selection flag, and the origin of the source (L10 or \citealt{best12}, B12). 
Choosing sources with the selection flag with value of 1 allows one to obtain the $z\le 0.15$ RG sample used  in Section~\ref{sec:halomass} (further restricting the redshift to $z\le 0.092$ gives the $z\le 0.092$ sub-sample).

We list in Table~\ref{tab:accept} the galaxies used in the analysis in Section~\ref{sec:icm}, providing the coordinates, cluster redshift, clustercentric distance, stellar mass, a flag indicating whether the galaxy is the BCG, radio luminosity (a value of $-1$ indicates luminosity below our threshold of $P_{\rm th}=23.5$), mean temperature of the cluster, local entropy, local pressure, local cooling time,  the ratio of local cooling time to free-fall time,  and the  name of the host cluster.

\begin{table*}
	\centering
	\caption{Radio galaxy sample used in Section~\ref{sec:halomass}}
	\label{tab:L10tab}
	\begin{tabular}{rrrrcccccc} 
		\hline
R.A. & Decl. & $z$ & $f_{1.4}$ & $\log P_{1.4}$ & $M_r^{0.1}-5\log h$ & $\log M_*$ & AGN? & Selection & Ref\\
(J2000) & (J2000) & & (mJy)   &  (W\,Hz$^{-1}$) &  & ($h^{-2}M_\odot$) & & & \\ \hline
\hline
2.814975 & $-9.272160$ & 0.21137 & 51.5 & 24.81 & $-22.844$ & 11.36 & 1 & 0 & L10\\
7.887342 & $-0.007844$ & 0.21945 &  6.7 & 23.96 & $-22.585$ & 11.25 & 1 & 0 & L10\\
6.942545 & $-10.541183$ & 0.16709 &  3.3 & 23.39 & $-22.711$ & 11.31 & 1 & 0 & L10\\
8.979727 & $0.356457$ & 0.25916 &  3.6 & 23.85 & $-22.746$ & 11.35 & 1 & 0 & L10\\
5.067143 & $0.079453$ & 0.21236 & 107.7 & 25.13 & $-22.722$ & 11.34 & 1 & 0 & L10\\
260.534205 & $30.693860$ & 0.05009 & 10.2 & 22.78 & $-21.409$ & 10.64 & 0 & 0 & B12\\
260.374455 & $30.645760$ & 0.09398 &  9.5 & 23.32 & $-21.239$ & 10.74 & 1 & 1 & B12\\
\hline
	\end{tabular}
\end{table*}

\begin{table*}
	\centering
	\caption{Radio galaxy sample used in Section~\ref{sec:icm}}
	\label{tab:accept}
	\begin{tabular}{rrrrrrrrrrrrr} 
		\hline
R.A. & Decl. & $z$ & distance & $\log M_*$ & BCG? & $\log P_{1.4}$ & $\overline{T}$ & $\log K$ & $\log p$ & $\log t_{\rm c}$ & $t_{\rm c}/t_{\rm ff}$ & Cluster \\
(J2000) & (J2000) & & (kpc) & ($h^{-2}M_\odot$) & & (W\,Hz$^{-1}$) & (keV) & (keV\,cm$^{2}$) & (dyne cm$^{-2}$) & (Gyr) &  & \\  \hline
 230.833830 & $8.618041$ & 0.0351 & 159.1 & 10.47 & 0 & -1.00 &  3.6 & $ 2.41$ & $-10.84$ & $ 1.3$ & 1.99 & Abell2063\\
 230.780880 & $8.528385$ & 0.0351 & 193.4 & 10.50 & 0 & -1.00 &  3.6 & $ 2.47$ & $-10.97$ & $ 1.4$ & 2.06 & Abell2063\\
 247.159330 & $39.551266$ & 0.0300 &  0.0 & 10.98 & 1 & 24.90 &  4.5 & $ 1.14$ & $-9.40$ & $-0.5$ & 1.99 & Abell2199 \\
 247.163080 & $39.552837$ & 0.0300 &  7.0 & 10.75 & 0 & -1.00 &  4.5 & $ 1.35$ & $-9.72$ & $-0.2$ & 1.57 & Abell2199 \\
 220.162610 & $3.469738$ & 0.0270 &  5.0 & 10.85 & 0 & -1.00 &  3.3 & $ 2.17$ & $-10.41$ & $ 0.8$ & 2.41 & MKW08\\
\hline
	\end{tabular}
\end{table*}


\end{document}